\documentclass[a4paper]{jpconf}
\usepackage{graphicx}
\usepackage{amssymb}
\usepackage{amsmath}

\usepackage{float}
\usepackage[caption=false]{subfig}
\usepackage{graphicx}
\usepackage{hyperref}

\begin{document}
\title{Controlling Physical Attributes in GAN-Accelerated Simulation of Electromagnetic Calorimeters}

\author{Luke de Oliveira 
} 
\address{Vai Technologies \& Lawrence Berkeley National Laboratory }
\ead{lukedeo@ldo.io}
\author{Michela Paganini 
}
\address{Yale University Department of Physics \& Lawrence Berkeley National Laboratory}
\ead{michela.paganini@yale.edu}
\author{Benjamin Nachman
}
\address{Lawrence Berkeley National Laboratory}
\ead{bnachman@cern.ch}

\begin{abstract}

High-precision modeling of subatomic particle interactions is critical for many fields within the physical sciences, such as nuclear physics and high energy particle physics. Most simulation pipelines in the sciences are computationally intensive -- in a variety of scientific fields, Generative Adversarial Networks have been suggested as a solution to speed up the forward component of simulation, with promising results. An important component of any simulation system for the sciences is the ability to condition on any number of physically meaningful latent characteristics that can effect the forward generation procedure. We introduce an auxiliary task to the training of a Generative Adversarial Network on particle showers in a multi-layer electromagnetic calorimeter, which allows our model to learn an attribute-aware conditioning mechanism.

\end{abstract}

\section{Introduction}

Modeling the interactions of particles with media is critical across many many physical sciences. Detailed simulation of particle collisions and subsequent interactions at the LHC experiments, along with simulating exact detector response, is very computationally expensive, requiring billions of CPU hours and roughly half of LHC computing resources~\cite{Flynn:2002240,dashboard,Bozzi:1984010}.

Recently, deep learning-based generative models including Generative Adversarial Networks~\cite{goodfellow2014generative} and Variational Auto-Encoders~\cite{VAE}  have been proposed and tested as a solution to significantly speed up scientific simulation in Oncology~\cite{oncology}, Cosmology~\cite{CosmoGAN1,CosmoGAN2,cosmogan3}, and High Energy Physics~\cite{calogan,deOliveira:2017pjk}, and many other basic science fields. We propose a simple extension to the CaloGAN model~\cite{calogan} allowing such a generation model to be conditioned on a vector of physically meaningful characteristics. We introduce a series of auxiliary tasks to encourage our model to learn $p(x\vert\xi)$, where $\xi$ is a vector of meaningful characteristics which should guide the generation procedure.

\section{The Dataset}
\label{sec:data}

The publicly available dataset from Ref.~\cite{dataset}, composed of 500,000 $e^+$, 500,000 $\pi^+$, and 400,000 $\gamma$ \textsc{Geant4}-generated showers, is used to construct and validate a proposed architecture for conditional modeling. 
The detector geometry present in the data consists of a cubic region along the $z$ direction of $V = 480$ mm$^3$ of an ATLAS-inspired EM calorimeter, at a distance of $z_0 = 288$ mm from (0,0,0). The volume is radially segmented into three layers of depth 90 mm, 347 mm, and 43 mm composed of alternating layers of lead and liquid Argon.

In this work, all \textsc{Geant4} coordinates native to the simulation are transformed into ATLAS coordinates for consistency with practical use-cases in experiments.

\section{Generative Adversarial Networks}
\label{sec:gan}

Generative Adversarial Networks (GAN) ~\cite{goodfellow2014generative} are a method to learna generative model by recasting the generation procedure as a minimax game between two actors that are parameterized by deep neural networks. A \emph{generator} network is tasked with building samples that are very similar to some target data distribution, and a \emph{discriminator} network (the adversary) is tasked with learning to distinguish real-looking samples from fake-looking samples.

Formally, assume we have a data distribution we wish to model, $x\sim p_{\mathrm{data}}(x)\in\mathcal{X}$. We construct a generator network $G$ that maps a $d$-dimensional latent prior $z\sim p_z(z)\in\mathbb{R}^d$ to synthetic samples $G:\mathbb{R}^d \longrightarrow\mathcal{X}$. The map $G$ implicitly defines a learned probability distribution from which we can sample, $p_G(x)$. In order to direct $p_G(x)$ towards the data distribution $p_{\mathrm{data}}(x)$, a discriminator network $D$ is tasked with taking a sample $x$ and classifying it with a label $y$ as originating from the data distribution ($y=1$) or from the implicit synthetic distribution ($y=0$), i.e., $D:\mathcal{X}\longrightarrow [0, 1]$. 

In the original formulation of GANs~\cite{goodfellow2014generative} a loss $ \mathcal{L}_{\mathsf{adv}}$ is constructed (Eqn.~\ref{eqn:gan_formulation}) to guide the learning towards towards equilibrium. When $G$ and $D$ are allowed to be drawn from the space of all continuous functions (i.e., they have infinite capacity), this system converges to a unique Nash Equilibrium~\cite{nashgan} in which the implicit distribution $p_G(x)$ exactly recovers the target data distribution $p_{\mathrm{data}}(x)$, and $D(\cdot)$ is $1/2$ everywhere~\cite{goodfellow2014generative}. It can be shown~\cite{goodfellow2014generative} that this procedure minimizes the Jensen-Shannon divergence between $p_G(x)$ and $p_{\mathrm{data}}(x)$, $\mathrm{JSD}(p_G(x) \parallel p_{\mathrm{data}}(x))$.

\begin{equation}
  \mathcal{L}_{\mathsf{adv}} = 
    \underbrace{
        \mathbb{E}_{z \sim p_z(z)}[\log(1-D(G(z)))]
    }_{\begin{subarray}{l} 
        \text{term associated with $D$ classifying}\\
        \text{a sample from $G$ as fake}
       \end{subarray}} + 
    \underbrace{
        \mathbb{E}_{x \sim p_{\mathrm{data}}(x)}[\log D(x)]
    }_{\begin{subarray}{l}
        \text{term associated with $D$}\\
        \text{perceiving a real sample as real}
       \end{subarray}}
\label{eqn:gan_formulation}
\end{equation}

In the original formulation, $G$ minimizes $\mathcal{L_{\mathsf{adv}}}$ while $D$ minimizes $-\mathcal{L_{\mathsf{adv}}}$, i.e., the game is zero sum. However, this formulation suffers from gradient saturation when a synthetic sample $x=G(z)$ is seen as very fake, i.e., when $D(G(z))\approx 0$, stagnating the learning procedure due to near-zero gradients. To overcome this, Ref.~\cite{goodfellow2014generative} proposes the non-saturating heuristic, replacing the generator objective with Eq.~\ref{eq:heuristic}, which it minimizes just as before. This formulation is utilized in our experiments, and we let $\mathcal{L}_D = \mathcal{L_{\mathsf{adv}}}$ represent the same quantity as in Eq.~\ref{eqn:gan_formulation}, except now it is only used for the discriminator.

\begin{equation}
\label{eq:heuristic}
\mathcal{L}_G = - \mathbb{E}_{z \sim p_z(z)}[\log D(G(z))]
\end{equation}

Through architecture design~\cite{dcgan,whatwheredraw,odena_acgan,stackgan,conditional_gan,improved_gan} and more theoretically-sound training objectives used~\cite{info_gan,wgan,improved_wgan,lsgan,fgan,softmaxgan,cramergan}, GANs have emerged as one of the most promising methods for neural networks to learn generative models of complicated and structured data spaces. We follow Ref.~\cite{evaluation} and impose task-specific metrics which allow us to move away from likelihood-level notions of quality which do not directly translate to tasks of interest. We utilize the original heuristic loss formulation (Eq.~\ref{eq:heuristic}) with label flipping~\cite{improved_gan} in lieu of mroe recent advances. The authors expect future work will experiment with effective conditioning mechanisms for other more popular GAN formulations such as the Wasserstein GAN~\cite{wgan} and the Cram\'{e}r GAN~\cite{cramergan}. 

To effectively condition on physical attributes, we add additional terms to the loss outlined in Eqn.~\ref{eqn:gan_formulation}, which we explore in-depth in Sec.~\ref{sec:model}. We modify the \textsc{CaloGAN} architecture~\cite{calogan} to perform the attribute-conditional task, leaning on inspiration from Ref.~\cite{odena_acgan} for conditioning mechanisms.

\section{The Conditional \textsc{CaloGAN}}
\label{sec:model}

To create a GAN based simulator that presents a useful solution for fast simulation, we need not only learn $p_{\mathsf{data}}(x)$, but we must also approximate $p_{\mathsf{data}}(x\vert\xi)$, where $\xi$ is a vector of conditioning attributes. Given the dataset outlined in Sec.~\ref{sec:data}, $\xi$ is chosen to be $\xi=(E, x_0, y_0, \theta, \phi)$, where $x_0$ and $y_0$ are the incident coordinates and $\theta$ and $\phi$ are incident angles. As per Ref.~\cite{calogan}, an absolute-deviance is defined over requested and reconstructed energy from a simulated collision, encouraging the GAN to learn the conditional distribution. In this work, a generalization is presented. For quantities that do not have a direct mathematical formula given a shower (i.e., $x_0, y_0, \theta$, and $\phi$), we build a separate head on the discriminator to learn a specific submodel $\Xi$ which focuses on learning to regress on said quantities given a shower as input. This estimate $\hat{\xi} = \Xi(x)$ is learned along with the normal alternating gradient step. To both $\mathcal{L}_G$ and $\mathcal{L}_D$ from Sec.~\ref{sec:gan}, $\mathcal{L}_{\hat{\xi}}$ (Eq.~\ref{eq:condition_mechanism}) is added such that both players seek to minimize the attribute reconstruction error. 

\begin{equation}
\label{eq:condition_mechanism}
\mathcal{L}_{\hat{\xi}} = \lambda^T \mathbb{E}_{x\sim p_{\mathsf{data}},\; \xi\sim p_{\xi}(\xi\vert x)}\left[\vert\xi - \Xi(x)\vert\right]
\end{equation}

Note that in Eq.~\ref{eq:condition_mechanism} a hyperparameter $\lambda$ is included to allow us to individually control the contribution for each attribute reconstruction task. Combining this with the energy conditioning loss $\mathcal{L}_E$ and associated weight $\lambda_E$ from the original CaloGAN formulation~\cite{calogan}, our final optimization problem consists of the generator minimizing $\mathcal{L}_G + \mathcal{L}_{\hat{\xi}} + \lambda_E \mathcal{L}_E$ and the discriminator minimizing $-\mathcal{L}_D + \mathcal{L}_{\hat{\xi}} + \lambda_E \mathcal{L}_E$.

\section{Sparsity Considerations}
\label{sec:sparsity}
As in most scientific applications, sparsity plays an important role in determining how viable a generates sample is. In the original \textsc{CaloGAN} work~\cite{calogan}, a raw sparsity (or occupancy) calculation is performed and fed as additional information to the discriminator. For an image $X\in\mathbb{R}^{m\times n}$, the sparsity is calculated as:

\begin{equation}
\label{eq:occupancy}
\mathsf{sparsity}(X) = \frac{1}{mn}\sum_{i<m}\sum_{j<n} \mathbb{I}[X_{i,j}\neq 0],
\end{equation}

Note that although this allows the discriminator to learn to reject generated samples that do not match the sparsity levels of real samples, this formulation does not allow for any gradient signal to propagate to the generator due to all-zero sub-derivatives of the indicator function $\mathbb{I}[\cdot]$. To ameliorate this deficiency, \emph{soft sparsity} is introduced -- a quantity that in the limit of tunable hyperparameters, converges to Eqn.~\ref{eq:occupancy}. In particular, we define this quantity as

\begin{equation}
\label{eq:softsparse}
\mathsf{softsparsity}(X) = \left(\frac{1}{nm}\right) \left\Vert \frac{\vert X\vert^{\alpha} + \vert X\vert^{-\beta}}{\vert X\vert^{\alpha} + \vert X\vert^{-\beta} + 1 }\right\Vert_1,
\end{equation} 

where $\alpha,\beta>0$, all matrix powers and $\vert\cdot\vert$ operators are assumed to act \emph{pointwise} on $X$, and $\Vert\cdot\Vert_1$ is the entry-wise 1-norm rather than the induced norm. 

Examining Eqn.~\ref{eq:softsparse}, we note that $\mathsf{softsparsity}:\mathbb{R}^{m\times n}\longrightarrow [0, 1]$, and that the limit behavior for Eqn.~\ref{eq:softsparse} is consistent with Eqn.~\ref{eq:occupancy}, i.e., 
\begin{equation}
\lim_{\alpha,\beta\rightarrow\infty} \mathsf{softsparsity}(X) = \frac{1}{mn}\sum_{i<m}\sum_{j<n} \mathbb{I}[X_{i,j}\neq 0].
\end{equation}

This information is added to the discriminator in raw form as well as a minibatch discrimination~\cite{improved_gan} step in order to allow our model to learn the correct distribution of sparsity.

\section{Experiment \& Results}

\subsection{Setup}
A Conditional \textsc{CaloGAN} (Sec.~\ref{sec:model}) with soft sparsity information (Sec.~\ref{sec:sparsity}) is trained on the dataset described in Section~\ref{sec:data}. We perform our experiments on the simulated $e^{+}$ showers. 

To allow the model to learn to generate translation invariant samples, we replace all locally connected layers in the original \textsc{CaloGAN} formulation with normal convolution layers, as well as a fully-convolutional attention mechanism to mediate layer-to-layer dependence. In addition, convolutional kernels are added as blurring operations to specific detector layers for smearing effects, where weights are initialized to $\frac{1}{nm}$ for an $n\times m$ filter and are not updated during training, which was empirically found to improve generation quality. This explicit blur operation allows the model to learn a factored problem which marginalizes out smearing effects, and allows the model to focus on important factors in shower patterns.

Network trainings are performed using the \textsc{Keras}~\cite{keras} library with customized TensorFlow~\cite{tensorflow2015-whitepaper} operations across four Pascal Architecture NVIDIA\textsuperscript{\textregistered} Titan X GPUs for a total of 150 epochs.

\subsection{Results}

\begin{figure}%
\centering
\subfloat{
  \centering
      \includegraphics[width=0.18\textwidth]{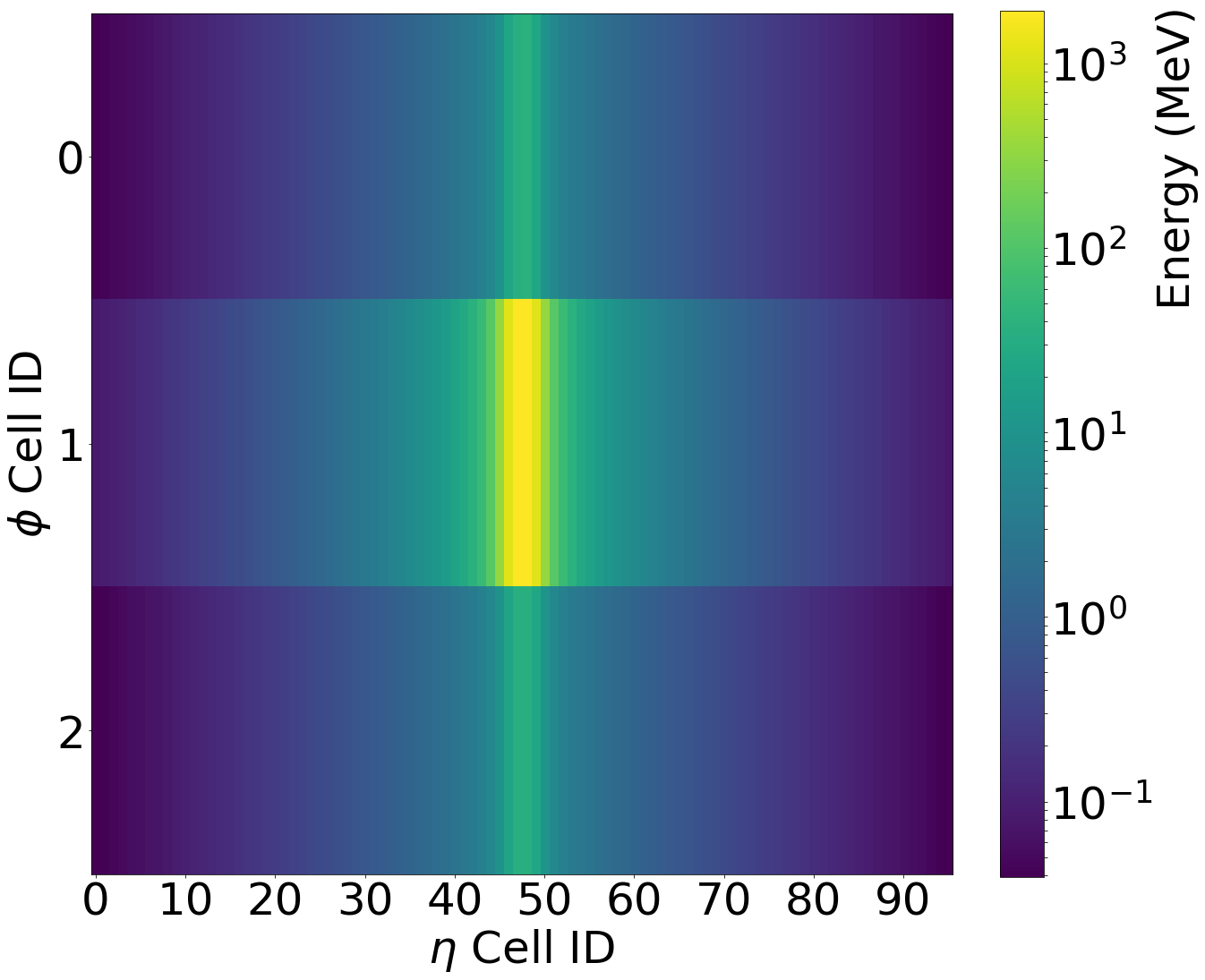}\hfill
      \includegraphics[width=0.18\textwidth]{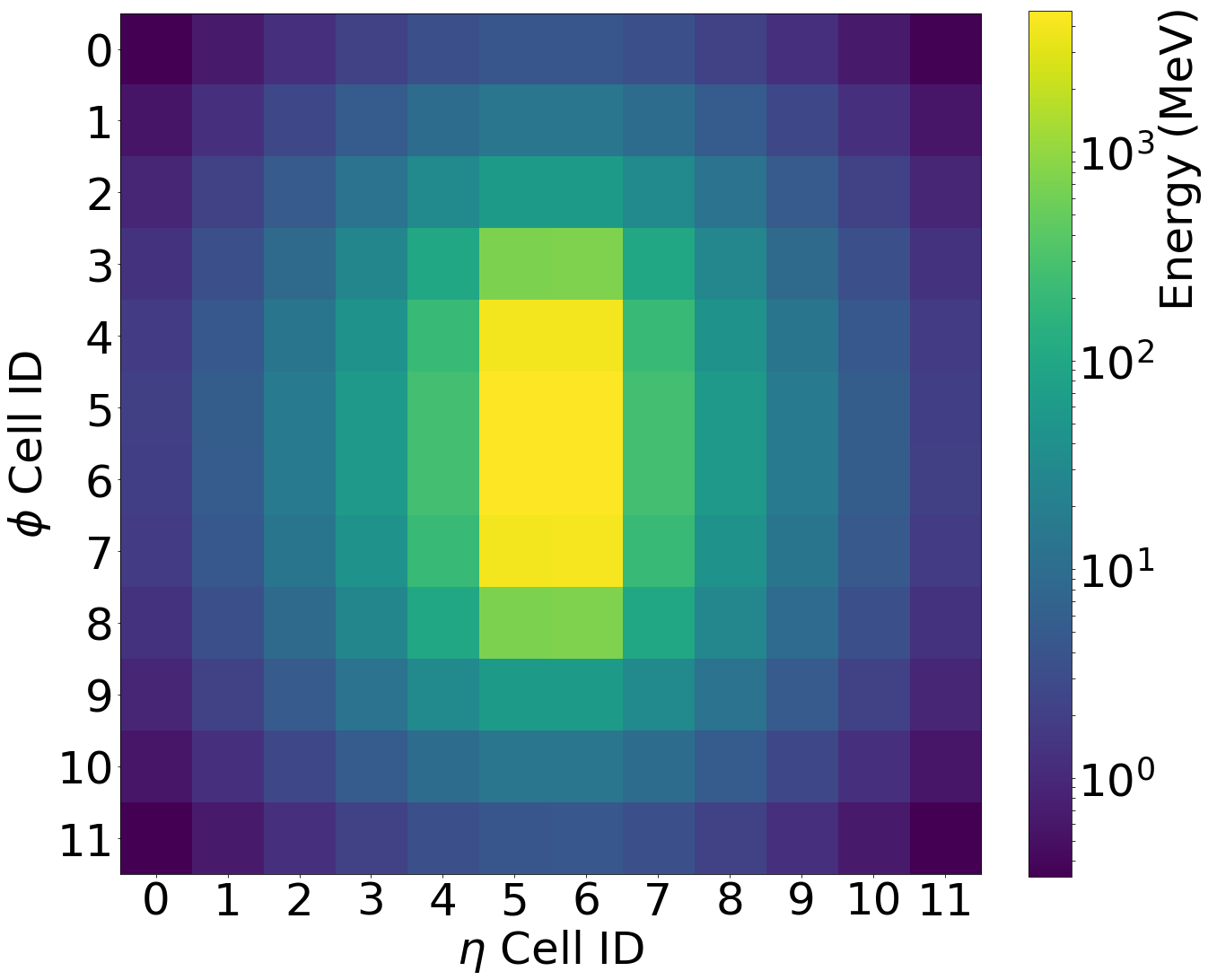}\hfill
      \includegraphics[width=0.18\textwidth]{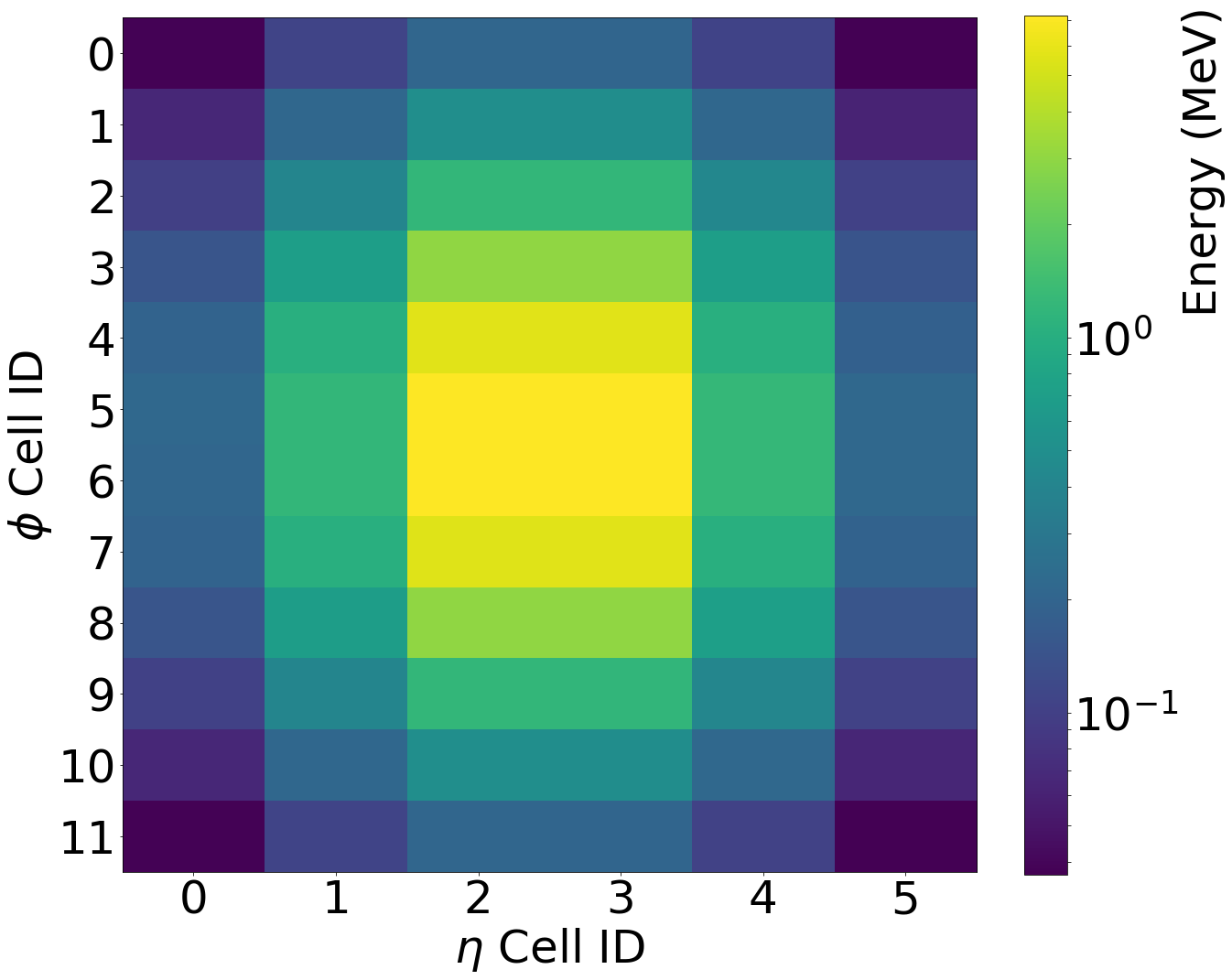}
}
\vspace{2mm}
\subfloat{
\centering
      \includegraphics[width=0.18\textwidth]{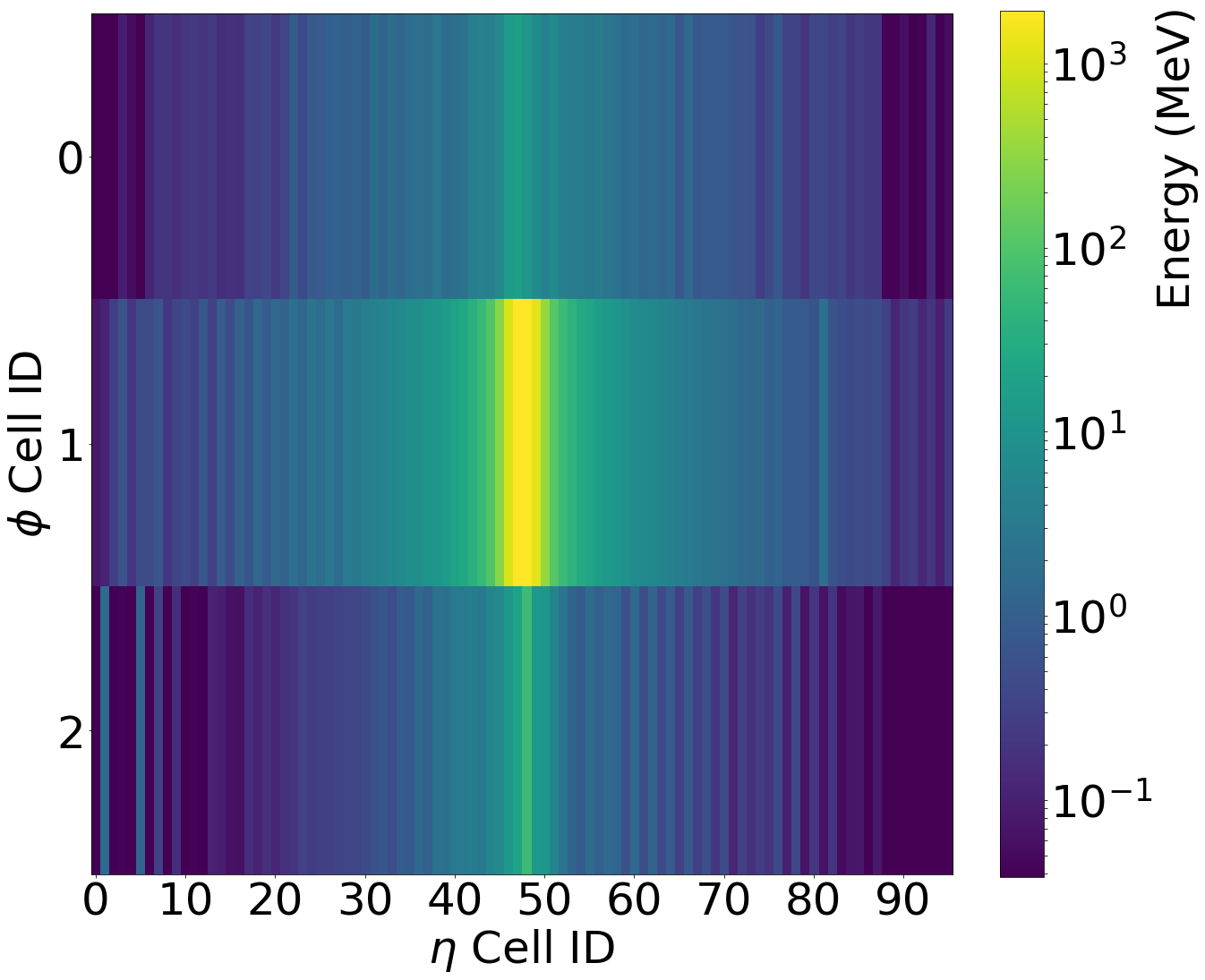}\hfill
      \includegraphics[width=0.18\textwidth]{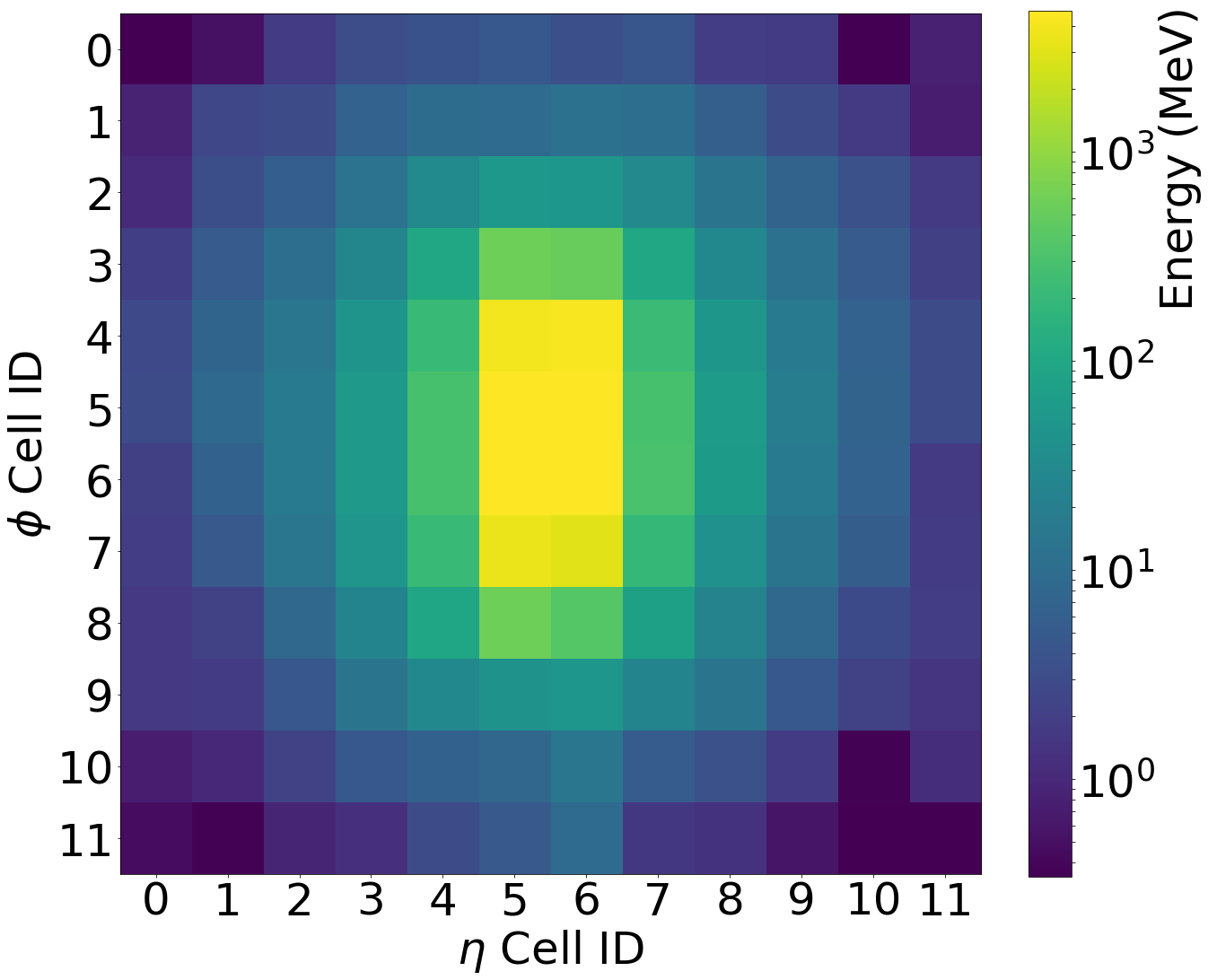}\hfill
      \includegraphics[width=0.18\textwidth]{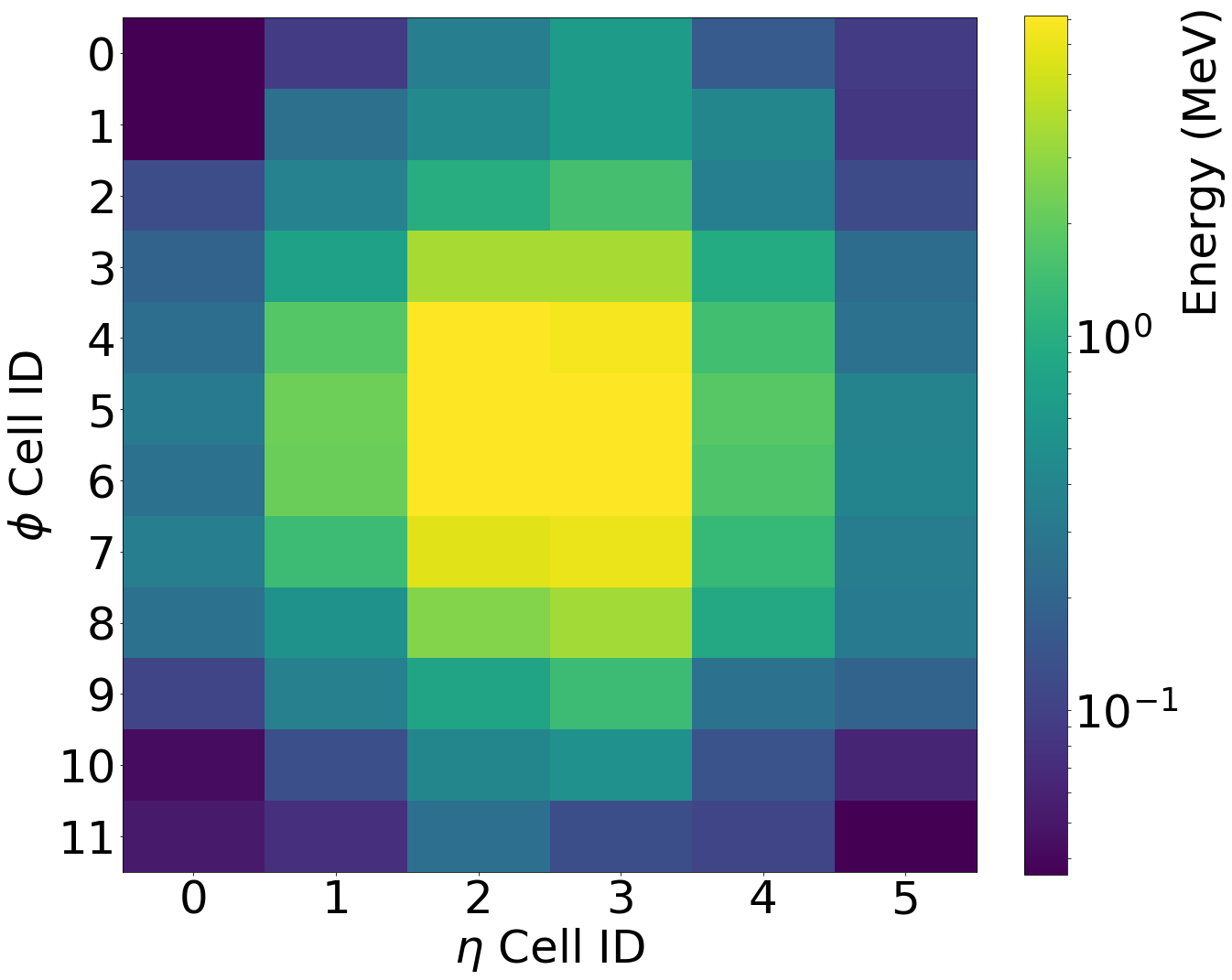}
}
\vspace{-3mm}
\caption{Average generated calorimeter image from \textsc{Geant} (top) and our model (bottom) for $e^{+}$. From left to right, we proceed through layers of the segmented calorimeter.}
\label{fig:avgs}
\end{figure}

As a first layer of validation, and commensurate with the validation performed in Ref.~\cite{calogan}, average images per calorimeter layer between images from the dataset of \textsc{Geant4} images are compared to images generated from the proposed model, as shown in Figure~\ref{fig:avgs}. This visualization illustrates the increased energy dispersion compared to the original dataset presented in Ref.~\cite{calogan}. The comparison in Figure~\ref{fig:avgs} shows a high level of visual agreement between GAN-generated and \textsc{Geant4}-generated showers.

In addition to aggregate image pattern validation, nearest GAN neighbors\footnote{By standard vectoral 2-norm.} are retrieved for seven \textsc{Geant4} images and used to validate that (a) our model does not memorize shower patterns, and (b) that the full space of displacements (both angular and positional) are explored.

\begin{figure}%
\centering
      \includegraphics[width=0.3\textwidth]{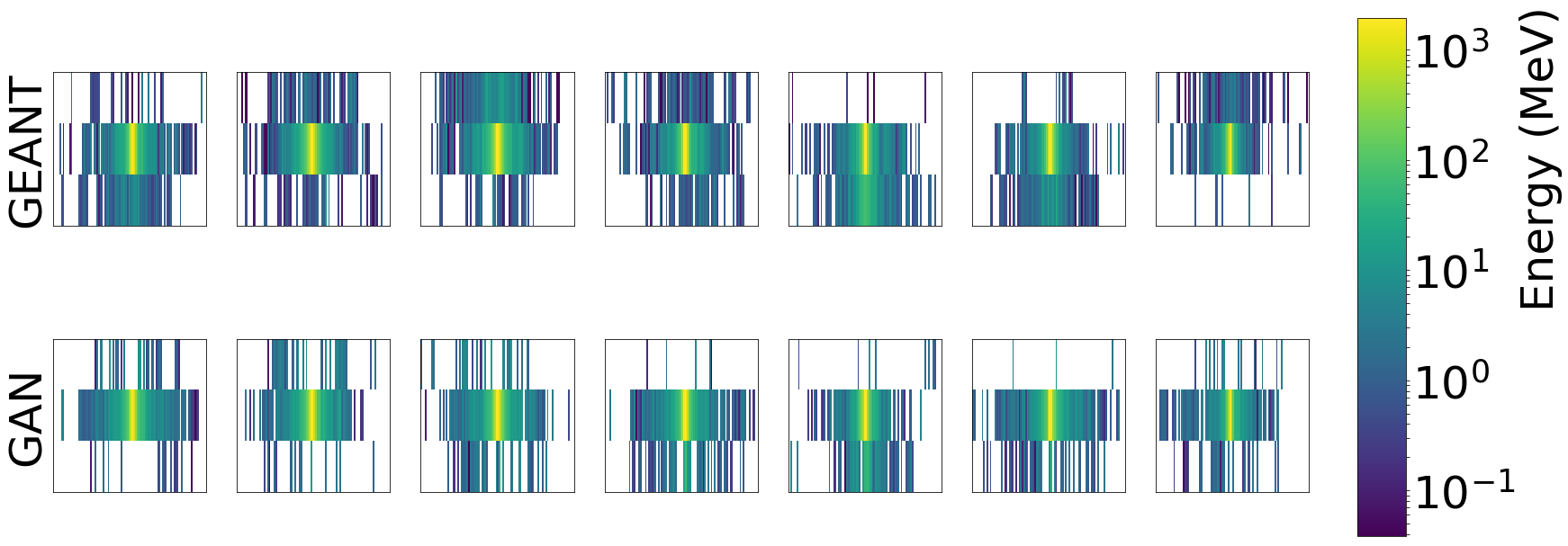}\hfill
      \includegraphics[width=0.3\textwidth]{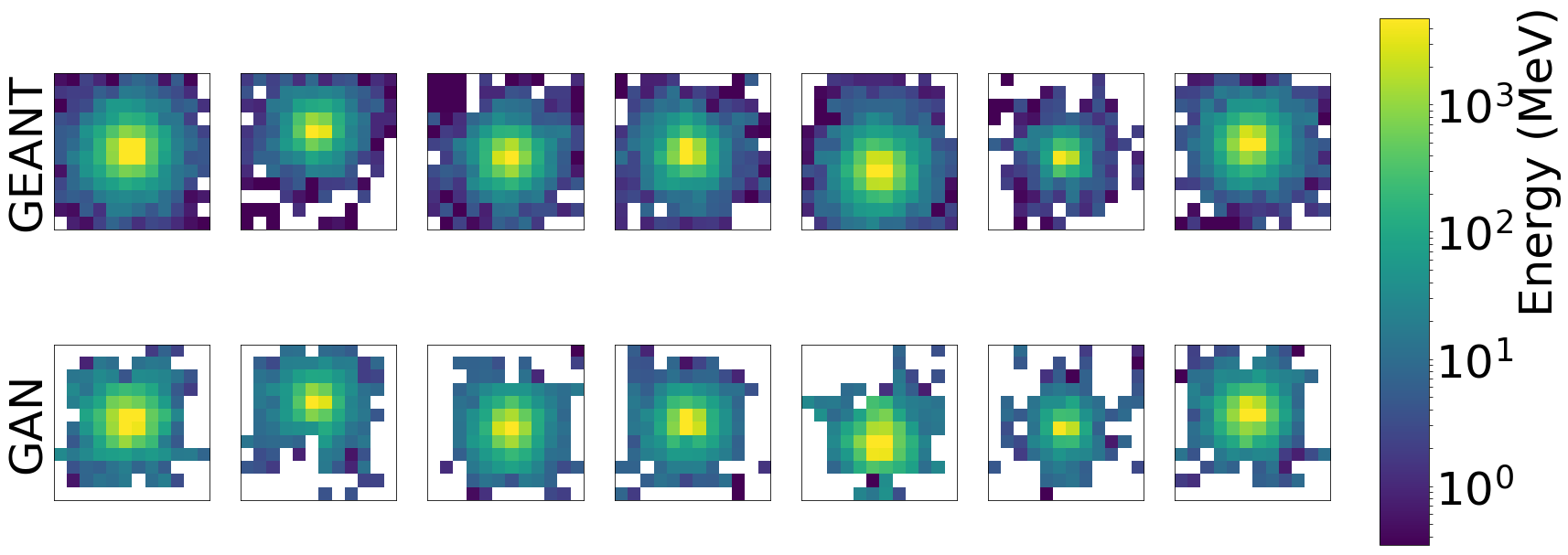}\hfill
      \includegraphics[width=0.3\textwidth]{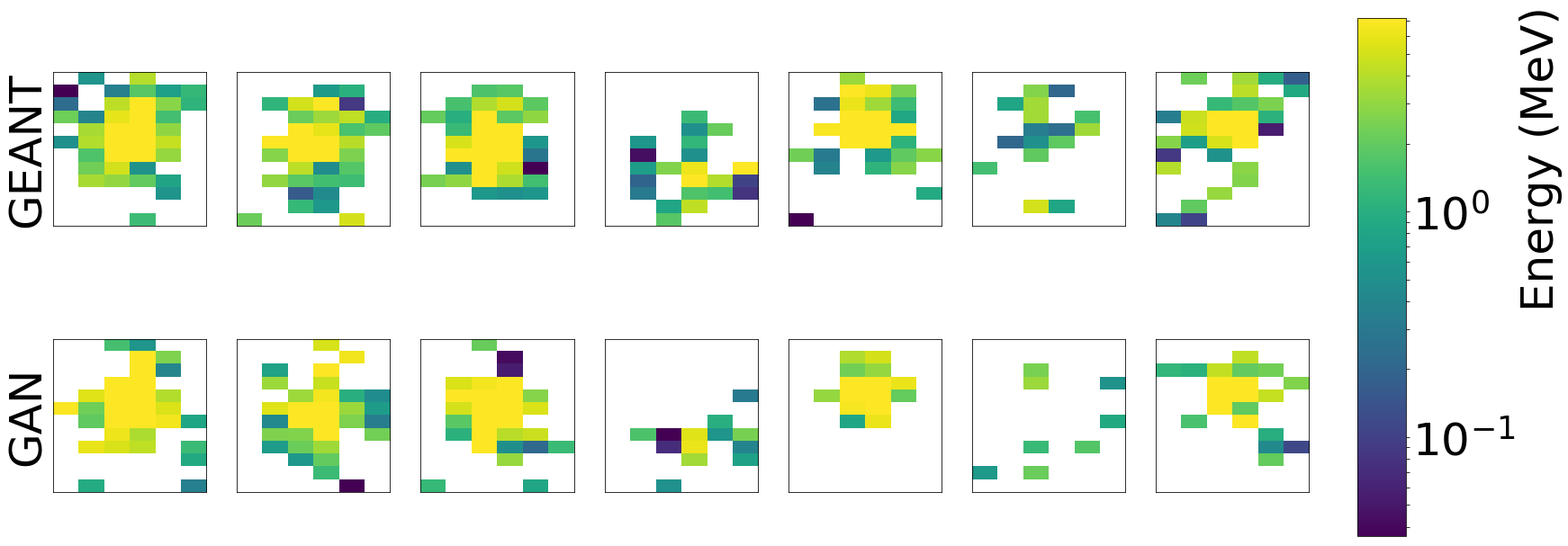}
\caption{Nearest GAN-generated neighbors (bottom) for seven random \textsc{Geant4}-generated $e^{+}$ showers (bottom) for the first layer (left), second layer (middle), and last layer (right) of the calorimeter.}
\vspace{-4mm}
\label{fig:nearest_neighbors}
\end{figure}

At the nearest-neighbor level, the model produces convincing energy deposition patterns, as shown in Figure~\ref{fig:nearest_neighbors}. The model does not appear to memorize the training dataset. In addition, positional variance (observed by noticing energy centroid deviations from the center of the calorimeter image) is well explored by the GAN, as shown by GAN-generated images matching all positions given by \textsc{Geant4}.

To further verify our models ability to condition on physical attributes, the latent space for each conditioning variable is traversed, showing how the model learns about each conditioning factor. In any practical setting, such conditioning mechanisms will need to be tuned to a high level of fidelity.

\begin{figure}%
\centering
\includegraphics[width=0.7\textwidth]{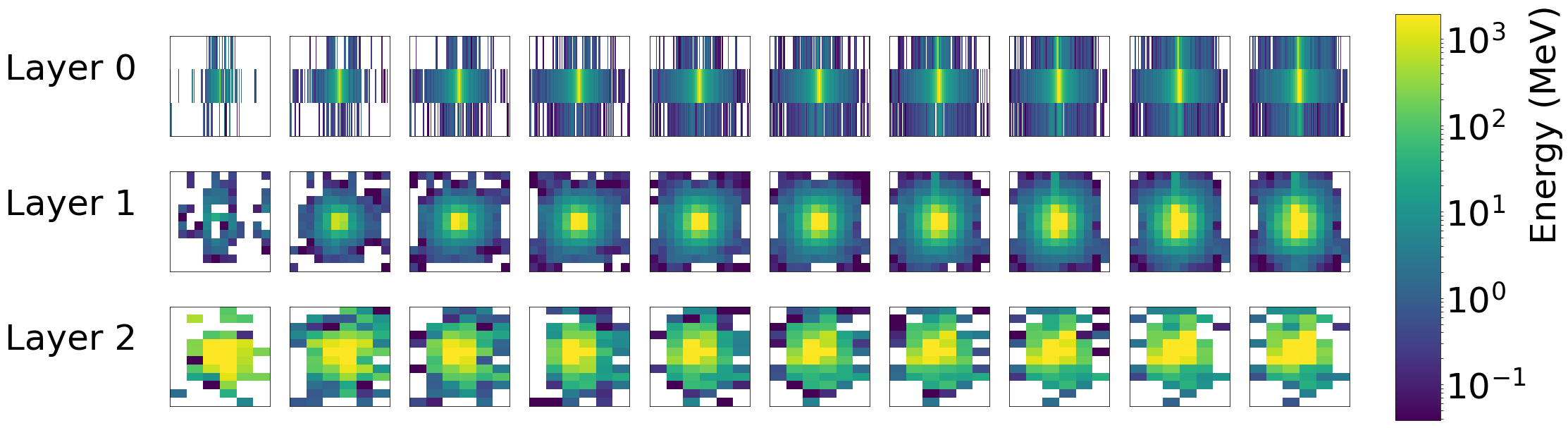}
\caption{Interpolation across physical range of incident energy as a conditioning latent factor for $e^{+}$ showers, with energy increasing from 1 GeV to 100 GeV from left to right. Each point in the interpolation is an average of 10 showers, with each point along the traversal build from an identical latent prior $z$.}
\label{fig:E_cond}
\end{figure}

To illustrate the model's internal representation, incident energy, $x_0$, and $\theta$ manifolds are traversed at regular intervals along the trained range. In Figure~\ref{fig:E_cond}, incident energy is traversed, clearly showing more energetic behavior as the incident energy is increased from left to right. 

\begin{figure}%
\centering
\includegraphics[width=0.7\textwidth]{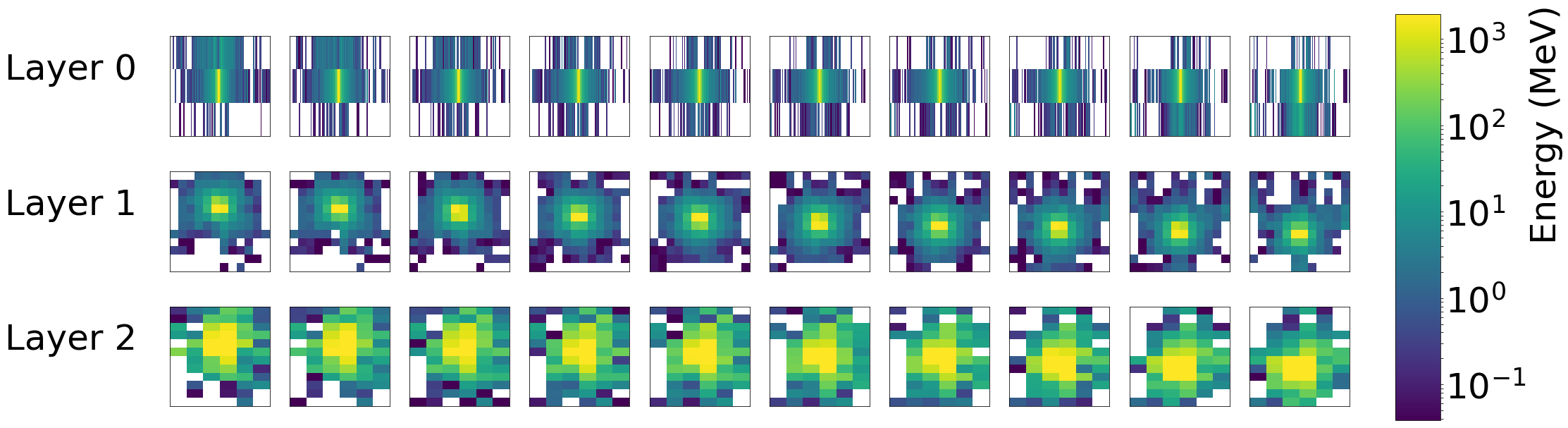}
\caption{Interpolation across physical range of $x_0$ as a conditioning latent factor for $e^{+}$ showers. Note in the ATLAS coordinate system, $x$ represents the vertical direction in this dataset. Each point in the interpolation is an average of 10 showers, with each point along the traversal build from an identical latent prior $z$.}
\label{fig:x0_cond}
\end{figure}

Similarly, the latent space for $x_0$ is traversed, and the resulting impact on generated image is shown in Figure~\ref{fig:x0_cond}. We note that as $x_0$ increases, shower position shifts downward, which is consistent with the ATLAS coordinates used in the dataset described in Sec.~\ref{sec:data}. 

\begin{figure}%
\centering
\includegraphics[width=0.7\textwidth]{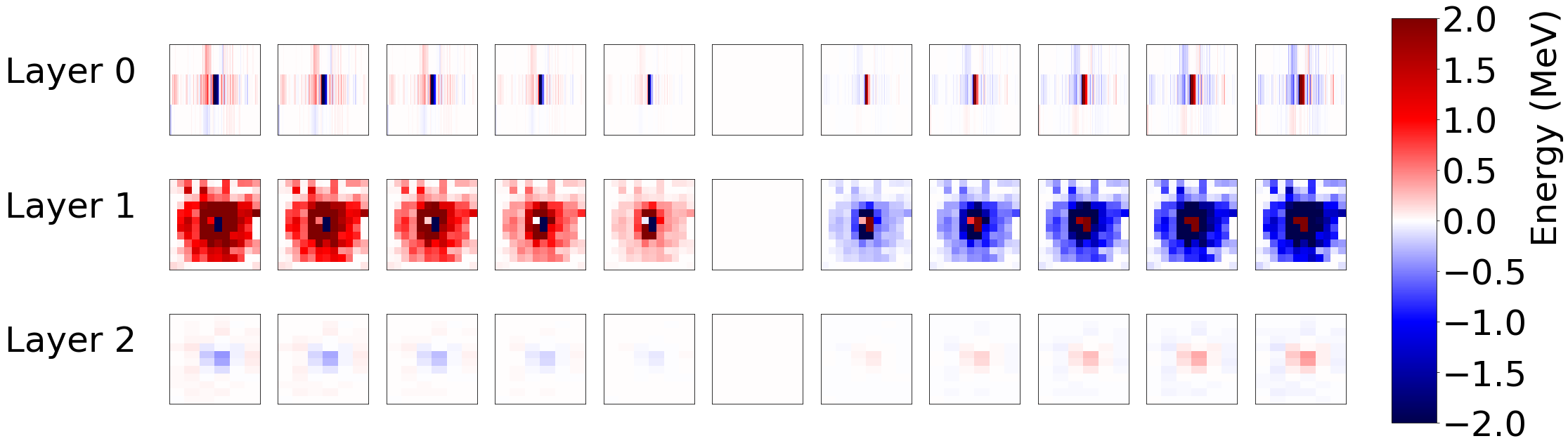}
\caption{Interpolation across physical range of $\theta$ as a conditioning latent factor for $e+$ showers, with $\theta$ increasing from left to right. Each point in the interpolation is an average of 10 showers subtracted from the middle point along the interpolation path, with each point along the traversal build from an identical latent prior $z$.}
\label{fig:theta_cond}
\end{figure}

Finally, as we traverse $\theta$ (Fig.~\ref{fig:theta_cond}) we illustrate the shower behavior dynamic using a difference between the middle point in interpolation space and each point along the $\theta$ traversal. As $\theta$ increases, we note that the width and dispersion decreases and the showers become significantly more centralized\footnote{In Figure~\ref{fig:theta_cond}, areas turning blue indicate that less energy is deposited in that particular section of the image at a given point in latent space.}, which is consistent with the ATLAS definition of $\theta$.

\section{Conclusion}

In this work, we explore the ability of GANs to be conditioned on physically meaningful attributes towards the ultimate goal of creating a viable, comprehensive solution for fast, high fidelity simulation of electromagnetic calorimeters. Clearly, GANs show great potential for controllability of generation procedures, but much future work remains. In particular, a thorough investigation around dynamics between the attribute estimation portion of the network, $\Xi$, and the overall training objective should be pursued, particularly as it relates to the final fidelity of the attribute estimates. In addition, future work should examine newer GAN formulations (as outlined in Sec.~\ref{sec:gan}) and their ability to improve image quality.

\vspace{20mm}

\bibliographystyle{iopart-num}
\bibliography{sample}

\end{document}